\documentstyle[10pt,aas2pp4]{article}

\def\micron{\,$\mu$m }
\def\etal{{et al.\,\,}}

\def\sec{\prime\prime}

\def\halfa{${\rm H\alpha}$ }
\def\Hone{H{\sc i} }
\def\Htwo{H{\sc i\sc i} }
\def\farcm{\hbox{$\> .\mkern-4mu^\prime$}}
\def\farcs{\hbox{$\> .\!\!^{\prime\prime}$}}

\begin{document}

\title{Optical Spectroscopy of Galactic Cirrus Clouds\footnote
{Observations carried out at the University of
California's Lick Observatory, Mount Hamilton, California}: 
Extended Red Emission in the Diffuse Interstellar Medium}

\author{Arpad Szomoru}
\and
\author{Puragra Guhathakurta\altaffilmark{2}}
\affil{UCO/Lick Observatory, University of California, Santa Cruz, California 95064}
\authoremail{arpad@ucolick.org}
\authoremail{raja@ucolick.org}
\altaffiltext{2}{Alfred P. Sloan Research Fellow}

\begin{abstract}

We present initial results from the first optical spectroscopic survey
of high latitude Galactic cirrus clouds.  The observed shape of the
cirrus spectrum does not agree with that of scattered ambient Galactic
starlight.  This mismatch can be explained by the presence of Extended
Red Emission (ERE) in the diffuse interstellar medium, as found in many
other astronomical objects, probably caused by photoluminescence of
hydrocarbons.  The integrated ERE intensity, $I_{\rm ERE}\approx
1.2\times 10^{-5}$\,erg\,s$^{-1}$\,cm$^{-2}$\,sr$^{-1}$, is roughly a
third of the scattered light intensity, consistent with recent color
measurements of diffuse Galactic light.  The peak of the cirrus ERE
($\lambda_{0}\sim 6000$\,\AA) is shifted towards short (bluer)
wavelengths compared to the ERE in sources excited by intense
ultraviolet radiation, such as \Htwo regions ($\lambda_{0}\sim
8000$\,\AA); such a trend is seen in laboratory experiments on
hydrogenated amorphous carbon films.

\end{abstract}


\section{Introduction}


A diffuse component to the Galactic interstellar medium (ISM) was
noticed over forty years ago \markcite{devau55}\markcite{devau60}(de
Vaucouleurs 1955, 1960) in the form of scattered starlight from dust
clouds \markcite{sandage76}(Sandage 1976).  It was the IRAS all-sky
survey though that first revealed cirrus as a ubiquitous component of
the far-infrared sky \markcite{low84} (Low \etal 1984), demonstrating
that these tenuous clouds of warm dust are representative of the diffuse
ISM in our Galaxy (\markcite{beich87}Beichman 1987;
\markcite{boul94}Boulanger 1994) and probably in all late-type spirals. 
\markcite{low84}Low \etal (1984) noted the positional correlation of the
cirrus emission with \Hone clouds mapped by \markcite{heiles75}Heiles
(1975); later optical studies of cirrus (\markcite{devries85}de Vries \&
le Poole 1985; \markcite{paley90}Paley 1990; \markcite{stark93}Stark
1993) revealed a good correspondence between optical and infrared
cirrus. 

Since the IRAS mission, analyses of the cirrus have shown that the
far-infrared emission from high latitude cirrus arises from an ensemble
of silicate and graphite dust grains and complex molecules heated by the
ambient interstellar radiation field (ISRF) (\markcite{low84}Low \etal
1984; \markcite{mezger82} Mezger \etal 1982).  The color temperature of
the cirrus emission at 12 and 25$\mu$m, where visible, indicates thermal
fluctuations of very small grains and/or polycyclic aromatic
hydrocarbons (PAHs) within clouds (cf.  \markcite{puget85}Puget \etal
1985; \markcite{boul85} \markcite{boul88a}Boulanger \etal 1985, 1988;
\markcite{heiles88}Heiles \etal 1988; \markcite{guha89b}Guhathakurta \&
Draine 1989).  There is at least an order of magnitude variation in the
infrared colors from cloud to cloud, indicating considerable
compositional variation.


In addition to scattered and thermally reprocessed radiation, Extended
Red Emission (ERE), a broad emission feature of $\sim1000$\,\AA\
width centered between 6000\,\AA\ and 8000\,\AA, has been detected in
the spectra of a wide variety of Galactic objects: reflection nebulae
\markcite{schmidt80}\markcite{witt88}(Schmidt \etal 1980; Witt \& Schild
1988), a high latitude Galactic dark cloud
\markcite{mattila79}\markcite{chlewicki87}(Mattila 1979; Chlewicki \&
Laureijs 1987), planetary nebulae (PN)
\markcite{furton90}\markcite{furton92}(Furton \& Witt 1990, 1992), H{\sc
ii} regions \markcite{perrin92}\markcite{sivan93} (Perrin \& Sivan 1992;
Sivan \& Perrin 1993), and a nova \markcite{scott94}(Scott \etal 1994). 
Recently, ERE has also been detected outside the Galaxy, in the halo of
M82 \markcite{perrin95}(Perrin \etal 1995). 

Possible carriers of the ERE include hydrogenated amorphous carbon
(HAC--- \markcite{duley85}Duley 1985;
\markcite{duley88}\markcite{duley90} Duley \& Williams 1988, 1990), PAHs
(\markcite{hendec86}d'Hendecourt \etal\\ 1986), and C$_{60}$
(\markcite{webster93}Webster 1993).  There is no obvious ``best''
candidate.  PAHs have been identified with the so-called unidentified
infrared emission (UIR) features and indeed, in the reflection nebula
NGC\,2023, emission at 3.3\micron is correlated spatially with ERE
(\markcite{burton88}Burton \etal 1988; \markcite{witt89} Witt \& Malin
1989).  However, \markcite{perrin92}Perrin \& Sivan (1992) do not find
such a correlation in the Orion nebula; moreover, 3.3\micron emission
has been detected in PN with no ERE \markcite{furton92}(Furton \& Witt
1992).  Laboratory spectra of C$_{60}$ compare well with the observed
ERE in the PN NGC\,2327 \markcite{webster93}(Webster 1993).  The peak
and shape of the ERE feature though vary for different objects, and it
is unclear whether C$_{60}$ can reproduce this variability.  Whether ERE
is present in all dust or only in discrete dusty objects remains, to
some degree, an open question.  The study of ERE in the warm diffuse ISM
has important implications for the chemistry of and energy balance in
this phase. 


There is indirect evidence of ERE in the Galaxy's warm diffuse
interstellar medium.  Unusually red broadband colors ($B-R$, $R-I$) have
been measured for cirrus clouds (\markcite{guha89a}Guhathakurta \& Tyson
1989; \markcite{paley91}Paley \etal 1991; \markcite{guha94} Guhathakurta
\& Cutri 1994), suggesting the presence of an additional component (over
scattered light) in the $R$- and $I$-bands.  Recent work by
\markcite{gordon97}Gordon \etal (1997) reached a similar conclusion
using a very different technique: they decomposed Pioneer 10 \& 11 blue
and red surface brightness maps into contributions of Galactic starlight
and diffuse Galactic light.  The advantage of this approach is the wide
field of view, while the obvious disadvantage is poor angular
resolution.  All of these studies lack the spectral resolution needed to
make an unambiguous detection of the ERE bump.  Their interpretation is
further complicated by the lack of detailed knowledge of the color of
incident starlight and the amount of internal reddening within clouds. 

This paper presents the first direct spectroscopy of diffuse, high latitude
cirrus clouds illuminated by ambient Galactic starlight.  The targets of
study have low optical surface brightness, only $\sim1$\% of the night sky
brightness (they are not bright enough to be classified as reflection
nebulae), but the experiment is made possible by the availability of high
efficiency modern spectrographs.  These cirrus clouds have optical depths
substantially below unity (typically a few tenths) making them qualitatively
different from the previously studied dark cloud, L1780 \markcite{mattila79}
(Mattila 1979), with $A_B\sim2$--5~mag.  Our spectroscopic survey covers the
relatively tenuous outer edges of cirrus clouds, but even their densest
central portions show none of the usual signs of high opacity:  reduced
surface density of background stars or increased IRAS 100\,$\mu$m/60\,$\mu$m
ratio.  Sec.~2 contains a description of the observations and data reduction
techniques.  In Sec.~3, the spectra are analyzed in the context of scattering
of starlight and photoluminescence from hydrocarbons.  The main findings of
this paper are summarized in Sec.~4.

\section{Observations \& Data Reduction}


Our targets of study are compact, relatively isolated, high latitude
cirrus clouds selected from IRAS 100\micron maps or drawn from the
catalog of Lynds Bright Nebulae \markcite{lynds65} (Lynds 1965) which
is, in turn, derived from POSS plates.  Clouds with sharp edges are
ideal for making (differential) measurements of clouds relative to the
adjacent sky background.  The Lick Observatory Shane 3-meter telescope
and KAST spectrograph were used to observe three cirrus clouds -- Witch
(10$^{\rm h}$44$^{\rm m}$42$^{\rm s}$, +83$^{\rm o}43\farcm7$), LBN\,8
(15$^{\rm h}$50$^{\rm m}$3$^{\rm s}$, --5$^{\rm o}50\farcm2$), and Draco
(16$^{\rm h}$47$^{\rm m}$34$^{\rm s}$, +60$^{\rm o}17\farcm0$) -- during
2~nights in April~1997.  The KAST spectrograph's 300~lines~mm$^{-1}$
grating and Reticon 1200$\times$400 CCD (of which 1200$\times$200 are
read out) yield a wavelength range of
$\Delta\lambda=4090$\,\AA--9590\,\AA\ at a dispersion of
4.6\,\AA~pixel$^{-1}$ with 27\micron pixels and a scale of about
$0\farcs8$\,pix\-el$^{-1}$.  The slit width and length of 3$^{\sec}\times
$145$^{\sec}$ result in an overall resolution of 18.4\,\AA\ (FWHM).  The
observations consist of 20~min exposures $\times$ 15 pointings for the
Witch and 7 pointings for each of LBN\,8 and Draco, with each pointing
chosen to have the slit straddle a sharp edge in the cirrus.  Dome flat
spectra were obtained after every change of pointing and/or of slit
position angle.  The spectrophotometric standard stars Feige~34 and
Feige~98 were observed several times during the observing run. 
Observing conditions were initially photometric but degraded during the
course of the first night; the second night was not photometric.  The
seeing FWHM was $\sim 2^{\sec}$ during the observations. 


As both zero level and dark current are negligible, we do not apply any
overscan, bias, or dark current subtractions.  For each spectrum, a
flat-field image is constructed by averaging the best matched dome flats
(nearest in terms of telescope pointing and slit position angle).  Due
to flexure of the instrument however, the fringe patterns in spectra and
flat fields do not correspond exactly.  To correct for this effect, we
allow for a small shift of the flat fields along the dispersion axis. 
The optimal shift, typically $\sim$ 1--2 pixels, is determined by
minimizing the rms brightness variation along the slit of several bright
night sky emission lines in the flat-fielded images.  The spectra are
then corrected for geometric distortion in the spectrograph optics by
rectifying the night sky lines.


\begin{figure}[h]
\plotfiddle{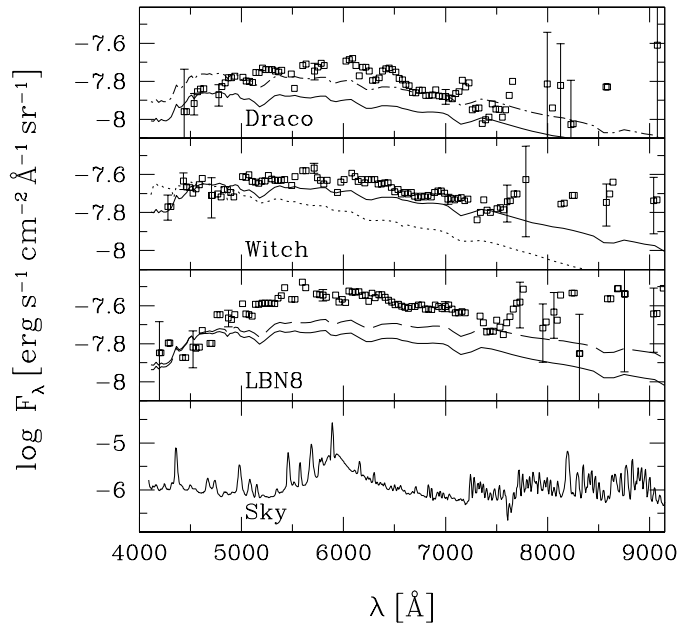}{11cm}{0}{58}{58}{-170}{-50}
\vspace{-2.5cm}
\caption{Spectra of three cirrus clouds, in order of
decreasing surface brightness.  Representative errorbars are shown,
spanning the smallest and largest errors at the blue and red ends of the
spectrum.  The solid lines are scattered synthetic E-type galaxy spectra
(Poggianti 1997) normalized to the cirrus spectra at $\sim 4500$\,\AA.
Normalizing the scattered light model at 5000\,\AA\ (dot-dashed line in first
panel) reduces the inferred amount of ERE but results in a deficit for
$\lambda<4800\,$\AA; the dotted line in the second panel is based on an
Sc-type ambient spectrum; the long-dashed line in the third panel corresponds
to twice the nominal optical depth.  The lowest panel shows the night sky
spectrum.}
\end{figure}

The low surface brightness of the cirrus ($B\sim26$\,mag\,arcsec$^{-2}$,
$R\sim24$\,mag\,arcsec$^{-2}$) makes this a challenging project, and
requires very accurate subtraction of the night sky spectrum, especially
its bright emission lines.  For a few of the spectra the task is
facilitated by the fact that the long slit extends to cirrus-free parts
of the sky on both sides of the cloud.  The other cases are more
complicated, and we use a multi-step iterative process: {\bf[1]}~A
relatively emission line free part of the night sky spectrum
(6350\,\AA--7515\,\AA) is collapsed in the dispersion direction
(combining several columns) to determine the exact location of the
cirrus edge ($l_{\rm edge}\approx0.5\,l_{\rm max}$), and the brightness
profile of the cirrus along the slit length, $I(l)$.  {\bf[2]}~The
cirrus free section [0, $l_{\rm edge}$] is used to make a preliminary
measurement of the sky spectrum, which is then subtracted from each row
of the full 2D spectrum.  {\bf[3]}~This extrapolation of the night sky
spectrum beyond $l>l_{\rm edge}$ has large uncertainties, however,
primarily because of imperfect rectification.  We isolate a small
($\Delta{l}\sim 10^{\sec}$) section of the (preliminary) sky-subtracted
spectrum at the end of the slit opposite the cirrus-free end,
[$0.9\,l_{\rm max},~l_{\rm max}$], and make a low-order polynomial fit,
$I_{\rm fit}^{\rm cirr}$, to the cirrus spectrum (relying on the
assumption that it is fairly smooth in $\lambda$).  {\bf[4]}~This
smooth fit, $I_{\rm fit}^{\rm cirr}$, is used to subtract off the
contribution of the cirrus from the [$0.9\,l_{\rm max},~l_{\rm max}$]
section of the ``original'' 2D spectrum (i.e.,~the pre-sky subtraction,
flat-fielded and rectified spectrum).  The final night sky spectrum is
then interpolated from two extreme sections of this residual 2D
spectrum: [$0,~l_{\rm edge}$] and [$0.9\,l_{\rm max},~l_{\rm max}$]. 
This final night sky spectrum is subtracted from each row of the 2D
spectrum. 


Next, each sky-subtracted spectrum is collapsed into a median cirrus
spectrum over the appropriate section of the slit length: [$l_{\rm
edge},~0.9\,l_{\rm max}$].  Median spectra from individual exposures of
a given cloud are then combined using a noise-weighted average, with the
noise determined over the ``line-free'' 6350\,\AA--7515\,\AA\ range. 
The rms scatter in each $\lambda$ pixel amongst individual spectra is
used to mask poorly subtracted night sky emission lines, especially at
the red end where residual fringes are a problem.  The resulting masked
spectrum is then boxcar-smoothed in $\lambda$, ignoring masked pixels, to a
resolution of 97\,\AA\ in order to enhance the signal-to-noise ratio. 

Wavelength calibration is done using night sky lines \markcite{oster92}
(Osterbrock \& Martel 1992) in a star- and cirrus-free part of one of
the spectra.  Initial flux calibration is based on observations of the
Feige~34 spectrophotometric standard star.  Second order overlap,
combined with the high flux of Feige~34 for $\lambda\leq 4000$\,\AA,
leads to calibration errors for $\lambda\geq7500$\,\AA.  To correct
this, we use the initially calibrated spectra of a number of field stars
that the long slit happened to intersect.  Most of these stars are
easily identifiable as K and M dwarfs, as their observed spectra closely
resemble empirical stellar library spectra \markcite{fioc97} (Fioc \&
Rocca-Volmerange 1997) for $\lambda\leq7500$\,\AA.  This enables us to
determine a residual flux calibration correction for $\lambda\geq
7500$\,\AA\ (averaged over all the field stars) which is then applied to
the cirrus spectra.

\section{Evidence for Extended Red Emission}


Figure~1 shows the calibrated spectra of all three cirrus clouds.  We
also plot a night sky spectrum to illustrate the strength of the night
sky lines, particularly for $\lambda > 7500$\,\AA.  Since no
particularly bright star is seen in the vicinity of the clouds, it is
reasonable to assume that the cirrus is illuminated by ambient Galactic
starlight.  This assumption is justified because: (1) the spectral
shapes of clouds in different locations are roughly similar, and (2) our
measurement represents an average over multiple slit positions within
each cloud, effectively averaging out the illumination effects of
individual bright stars.

We compare the observed spectral shape of the cirrus to a simple
scattering model.  The spectrum of the incident radiation field is
represented by synthetic galactic spectra of various Hubble types
\markcite{pogg97}(Poggianti 1997).  The optical depth estimate is based
on a standard conversion from IRAS 100\micron brightness
\markcite{boul88b} \markcite{savage79}(Savage \& Mathis 1979; Boulanger
\& Perault 1988): $A_{V}=0.053\,(S_{100}/1\,{\rm MJy\,sr^{-1}})$.  This
yields values of $A_{V} = 0.27$ for the Witch and Draco and $A_{V} =
0.64$ for LBN\,8.  In Fig.~1 we also plot the scattered starlight model
normalized to the cirrus emission at $\sim 4500$\,\AA, based on an incident
E-type galaxy spectrum.  The grain albedo is taken to be constant in
$\lambda$ at $\omega= 0.55$ \markcite{draine84}(Draine \& Lee 1984),
while the extinction law of \markcite{cardelli89}Cardelli \etal (1989)
is adopted. 

It is clear that the overall {\it shape} of the cirrus spectrum is not
fit by scattered starlight alone.  In particular, the cirrus spectra  
have too much tilt/curvature in the region 4000\,\AA--7000\,\AA\ to be 
fit by any plausible mix of stars (spectra of galaxy types later than E
produce even worse fits).  Scaling the model spectrum to match the red part
of the cirrus spectrum ($\lambda\ga5000\,$\AA) results in a deficit for
$\lambda\la4800\,$\AA\ (cf.~top panel of Fig.~1); we have instead chosen to
scale the model spectrum to the blue side of the cirrus as the resulting
broad region with excess emission (ERE) has a natural explanation.  It is not
quite clear whether this excess emission extends beyond
$\lambda\sim7500$\,\AA\ where measurement/calibration uncertainties are
large.  Incidentally, the shape of the cirrus spectrum agrees with
previous broadband observations of cirrus clouds (cf.   
\markcite{guha89a}Guhathakurta \& Tyson 1989), which showed both $B-R$
{\bf\it and} $R-I$ colors to be too red to be explained by scattered
Galactic light.

Note, the $S_{100}$ values adopted are conservative in that they have not had
any large scale background subtracted (due to foreground zodiacal light, for
example); for LBN\,8 in particular, the actual brightness of the cloud, and the
inferred $A_V$, could be as much as a factor of~2 lower.      
The conversion factor of~0.053 used to derive $A_V$ from $S_{100}$ is well
within the range found by \markcite{stark95}Stark (1995) for Galactic dust
clouds, based on photometry and counts of background stars.  As an extreme
example, we adopt a conversion factor of~0.1 for the densest cloud in our
sample, LBN\,8.  The resulting scattered light spectrum (long-dashed line in
third panel of Fig.~1) is understandably somewhat (internally) reddened
relative to that derived using the nominal optical depth (solid line), but
the conclusions about ERE in the cirrus are essentially unchanged.  The
effect of doubling the nominal optical depth is negligible for the other two
clouds.  Multiple scattering is ignored because $\tau_V^{\rm sca}<0.6$ even
for the densest cloud (LBN\,8) using an extreme $S_{100}\rightarrow{A}_V$
conversion ratio, whereas such events are only important if $\tau^{\rm
sca}>1$.

We also check whether the observed flux density of the cirrus is
consistent with an independent estimate of the strength of the ISRF in
the solar neighborhood (Mathis \etal \markcite{mathis83}(1983):
$4\pi\,J_{\lambda}\,(7000\,{\rm \AA})=1.5$ and
3.1\,erg\,s$^{-1}$cm$^{-2}\mu$m$^{-1}$, at galactocentric distances of
10 and 8\,kpc, respectively.  Applying the above dust scattering
parameters to the actual cirrus spectra we derive values for the
incident ISRF of 1--2\,erg\,s$^{-1}$cm$^{-2}\mu$m$^{-1}$. The
agreement with Mathis et al.'s estimate is remarkable when one considers
that: (1) it is difficult to measure the effective optical depth of the
cirrus averaged over the exact area represented by the spectrum, given
the poor angular resolution of the IRAS 100\micron maps; and (2) the
``isotropic'' ISRF strengths derived from the cirrus are likely to be
biased low if we are primarily seeing back scattered radiation from
preferentially forward scattering grains in clouds that are mostly
illuminated on one side by the Galactic disk.


The absence of a detectable scattered \halfa emission line in the cirrus
spectrum suggests that the radiation incident on the clouds has the
spectrum of an early type galaxy, E or Sa.  This is understandable if
the cirrus is local (heliocentric distance $\la300$~pc), in which case
it is illuminated mainly by a part of the disk between spiral arms
(which are separated by roughly 2--3~kpc).  The inter-arm region is
dominated by stars that are somewhat older than the global average over
the Galaxy, the latter including a contribution from young stars in star
forming/\Htwo regions.  This explanation for the lack of \halfa is indirectly
corroborated by independent evidence that the cirrus ERE is excited by a
UV-poor radiation field (see below).  Moreover, the presence of ERE tends to
dilute the reflected \halfa emission line, if any, in the cirrus spectra.
For completeness, we also present the scattered light spectrum for an ambient
Sc-type radiation field (dotted line in second panel of Fig.~1); it is a
worse fit to the data than scattered light based on an E-type spectrum.

\begin{figure}[h]
\plotfiddle{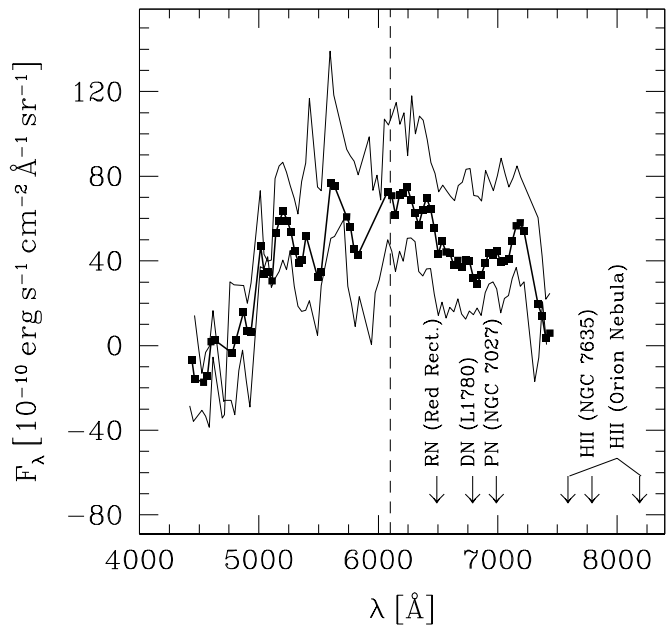}{12cm}{0}{58}{58}{-170}{-20}
\vspace{-3cm}
\caption{The difference between the cirrus spectrum
and the best fit (at $\sim 4500$\,\AA) scattered light model, averaged over
the three clouds (bold line) and for LBN\,8 and Witch (upper and lower thin
solid lines).  The dashed vertical line shows the approximate central
wavelength of the ERE. The arrows show the peak wavelengths of the ERE in a
number of astronomical objects; a reflection nebula (RN), a dark nebula (DN),
a planetary nebula (PN) and \Htwo regions.}
\end{figure}

In Fig.~2 we plot the average difference between the cirrus spectrum and the
best fit (at $\sim 4500$\,\AA) scattered light model, over the
4000\,\AA--7500\,\AA\ range within which flux measurements are reliable.
The thin solid lines indicate the ERE for LBN\,8 and Witch, the clouds with
the strongest and weakest excess, respectively.  These also provide some
measure of the uncertainty in our estimate of the average cirrus ERE 
spectrum.

Furton \& Witt \markcite{furton93}(1993) have conducted laboratory
experiments on HAC and AC (pure amorphous carbon) films to address
astronomical observations of ERE.  They find that UV radiation can
either diminish or enhance the efficiency of photoluminescence (PL) in
hydrogenated carbon solids, depending on the availability of free atomic
hydrogen and on the relative H saturation of the solid.  UV-annealing
and subsequent hydrogenation of HAC dust causes a broadening and shift
of the ERE peak towards the red.  Furton \& Witt note that the observed
shifting of the peak ERE emission to longer wavelengths in objects with
intense UV radiation, such as H{\sc i\sc i} regions and planetary
nebulae, must indeed be a consequence of this. Comparing the shape of the
cirrus ERE (Fig.~2) with laboratory PL yield (Fig.~1 of Furton \& Witt
\markcite{furton93}1993), we see excellent agreement in both shape and
width of the feature.  We also indicate in Fig.~2 the approximate
central wavelengths $\lambda_{0}$ of the ERE in a number of astronomical
objects: the H{\sc i\sc i} regions NGC~7635 \markcite{sivan93}(Sivan \&
Perrin 1993) and the Orion Nebula \markcite{perrin92}(Perrin \& Sivan
1992), the planetary nebula NGC~7027 \markcite{furton90}(Furton \& Witt
1990), the dark nebula L1780
\markcite{mattila79}\markcite{chlewicki87}(Mattila 1979; Chlewicki \&
Laureijs 1987), and the reflection nebula called the ``Red Rectangle''
around HD~44179 (\markcite{schmidt80}Schmidt \etal 1980).  In agreement
with Furton \& Witt's laboratory experiments, the peak of the cirrus
ERE, $\lambda_{0}\sim 6000$\,\AA, appears to be located at bluer
wavelengths than the ERE emission seen in environments with higher UV
radiation.   

The integrated ERE intensity of the cirrus is $1.2\times
10^{-5}$\,erg\,s$^{-1}$\,cm$^{-2}$\,sr$^{-1}$ and the ratio of ERE to
scattered intensity is $\sim$ 0.3, in good agreement with the findings
of \markcite{gordon97}Gordon \etal (1997): $I_{\rm ERE}\approx
10^{-5}$\,erg\,s$^{-1}$\\cm$^{-2}$\,sr$^{-1}$ and $I_{\rm ERE}/I_{\rm
sca}= 0.05$--2 in the diffuse ISM.  For other objects $I_{\rm
ERE}/I_{\rm sca}$ ranges from 0.01--0.68 in reflection nebulae to 0.2--0.6
in \Htwo regions.  Applying the same conversions as
\markcite{gordon97}Gordon \etal (1997), we find the cirrus has an ERE
photon efficiency of $\sim$ 10\%.

The shape of inferred cirrus ERE spectrum looks remarkably similar to
laboratory ERE spectra and to ERE in other astrophysical objects, and there
is a perfectly reasonable explanation for the shift of the cirrus ERE peak to
the blue relative to objects with hotter illuminating sources.  Nevertheless,
the cirrus ERE spectrum derived from this initial study should be treated as
preliminary (pending more thorough follow-up observations) in light of the
uncertainties in the data and in the scattered light model discussed above.

\section{Summary}

We present long slit spectra straddling sharp edges in three high
latitude Galactic cirrus clouds.  The cirrus is clearly detected in each
case.  The shape of the cirrus spectrum does not match the shape of
scattered ambient interstellar radiation, modeled using synthetic galaxy
spectra.  Assuming the blue end of the cirrus spectrum is dominated by
scattered light, there is an excess emission in the range
5000\,\AA--7000\,\AA, in good agreement with the Extended Red Emission
(ERE) bump seen in other astronomical objects.  The integrated ERE
intensity ($\sim 10^{-5}$\,erg\,s$^{-1}$\,cm$^{-2}$\,sr$^{-1}$) and
ratio between ERE and scattered light ($\sim0.3$) agree well with values
inferred from color measurements of the diffuse Galactic light
\markcite{gordon97}(Gordon \etal 1997).  The position of the peak ERE
emission in cirrus ($\sim6000$\,\AA), compared to objects such as H{\sc
i\sc i} regions and planetary nebulae, is consistent with a relatively
UV-poor ambient radiation field. 

\acknowledgements

The initial phases of this project were supported by a Faculty Research
grant at UCSC.  AS and PG are supported in part by NASA LTSA grant
NAG\,5-3232; PG is supported in part by an Alfred P.  Sloan Foundation
fellowship.  It is a pleasure to thank Roc Cutri for help with target
selection and other aspects of the cirrus project.  We also thank Dan 
Kelson for help with his Expector spectroscopic
analysis program, Karl Gordon for sharing unpublished results, and the
referee for useful comments.

\end{document}